\documentstyle[preprint,tighten,aps,epsf,amstex,graphicx,times,floats]{revtex}

\begin{document}

\thispagestyle{empty}

\marginparwidth 1.cm
\setlength{\hoffset}{-1cm}
\newcommand{\mpar}[1]{{\marginpar{\hbadness10000%
                      \sloppy\hfuzz10pt\boldmath\bf\footnotesize#1}}%
                      \typeout{marginpar: #1}\ignorespaces}
\def\mda{\mpar{\hfil$\downarrow$\hfil}\ignorespaces}
\def\mua{\mpar{\hfil$\uparrow$\hfil}\ignorespaces}
\def\mla{\marginpar[\boldmath\hfil$\rightarrow$\hfil]%
                   {\boldmath\hfil$\leftarrow $\hfil}%
                    \typeout{marginpar: $\leftrightarrow$}\ignorespaces}

\def\ba{\begin{eqnarray}}
\def\ea{\end{eqnarray}}
\def\bq{\begin{equation}}
\def\eq{\end{equation}}

\renewcommand{\abstractname}{Abstract}
\renewcommand{\figurename}{Figure}
\renewcommand{\refname}{Bibliography}

\newcommand{\eg}{{\it e.g.}\;}
\newcommand{\ie}{{\it i.e.}\;}
\newcommand{\etal}{{\it et al.}\;}
\newcommand{\ibid}{{\it ibid.}\;}

\newcommand{\mx}{M_{\rm SUSY}}
\newcommand{\pt}{p_{\rm T}}
\newcommand{\et}{E_{\rm T}}
\newcommand{\del}{\varepsilon}
\newcommand{\sla}[1]{/\!\!\!#1}
\newcommand{\fb}{\;{\rm fb}}
\newcommand{\gev}{\;{\rm GeV}}
\newcommand{\tev}{\;{\rm TeV}}
\newcommand{\abi}{\;{\rm ab}^{-1}}
\newcommand{\fbi}{\;{\rm fb}^{-1}}

\newcommand{\lsusy}{\lambda'_{\rm SUSY}}
\newcommand{\llq}{\lambda'_{\rm LQ}}
\newcommand{\msbar}{\overline{\rm MS}}

\newcommand{\SP}{\scriptscriptstyle}
\newcommand{\stl}{\tilde{t}_{\SP L}}
\newcommand{\str}{\tilde{t}_{\SP R}}
\newcommand{\ste}{\tilde{t}_1}
\newcommand{\stz}{\tilde{t}_2}
\newcommand{\st}{\tilde{t}}
\newcommand{\gt}{\tilde{g}}
\newcommand{\mse}{m_{\tilde{t}_{\SP 1}}}
\newcommand{\msz}{m_{\tilde{t}_{\SP 2}}}
\newcommand{\mst}{m_{\tilde{t}}}
\newcommand{\mg}{m_{\tilde{g}}}
\newcommand{\ms}{m_{\tilde{q}}}
\newcommand{\mt}{m_t}
\newcommand{\mheavy}{m_{\rm heavy}}

\preprint{
\font\fortssbx=cmssbx10 scaled \magstep2
\hbox to \hsize{
\hfill\vtop{\hbox{MADPH-02-1309}
            \hbox{CERN-TH/2002-344}
            \hbox{IFT-P.093/2002}
            \hbox{November 2002}        } }
}

\title{ 
          Stop Lepton Associated Production at Hadron Colliders
} 

\author{
Alexandre Alves${}^1$, Oscar \'Eboli${}^{2,4}$ and Tilman Plehn${}^{3,4}$ 
} 

\address{ 
${}^1$
Instituto de F\'\i sica Te\'orica, IFT-UNESP, S\~ao Paulo, Brazil \\
${}^2$
Departamento de F\'\i sica Matem\'atica, Universidade de S\~ao Paulo,
 S\~ao Paulo, Brazil \\
${}^3$
Physics Department, University of Wisconsin, Madison, USA \\
${}^4$
Theory Division, CERN, Geneva, Switzerland \\
} 

\maketitle 

\begin{abstract}

  At hadron colliders, the search for $R$-parity violating
  supersymmetry can probe scalar masses beyond what is covered by pair
  production processes. We evaluate the next-to-leading order SUSY-QCD
  corrections to the associated stop or sbottom production with a
  lepton through $R$-parity violating interactions. We show that
  higher order corrections render the theoretical predictions more
  stable with respect to variations of the renormalization and
  factorization scales and that the total cross section is enhanced by
  a factor up to $70$\% at the Tevatron and 50\% at the LHC. We
  investigate in detail how the heavy supersymmetric states decouple
  from the next-to-leading order process, which gives rise to a theory
  with an additional scalar leptoquark. In this scenario the inclusion
  of higher order QCD corrections increases the Tevatron reach on
  leptoquark masses by up to 40~GeV and the LHC reach by up to
  200~GeV.

\end{abstract}

\vspace{0.2in}

\section{Introduction}

Supersymmetry is one of the most promising candidates for physics
beyond the Standard Model, and the search for supersymmetric particles
is one of the most prominent tasks for current and future
colliders. Usually, searches for supersymmetric particles at colliders
assume the conservation of $R$ parity. However, exact $R$ parity is
not in any way inherent to supersymmetric models, neither gauge
invariance nor supersymmetry actually require it~\cite{expl,rp_coll}.
$R$ parity is imposed to bypass problems with flavor-changing neutral
currents, proton decay, atomic parity violation, and other
experimental constraints. It is also crucial for supersymmetric cold
dark matter. $R$-parity conservation has a huge impact on searches for
supersymmetric particles at colliders: superpartners can only be
produced in pairs and the final state has to include two stable
LSPs. In the absence of $R$-parity conservation, single supersymmetric
particles can be produced, which can extend the reach of
colliders~\cite{single_slepton,single_stop,single_susy_nlo}. It is
interesting to notice that pair production of scalar top quarks (or
sbottoms or scalar leptoquarks) in hadronic colliders is essentially
model independent since the squark--gluon interaction is fixed by
$SU(3)$ gauge invariance~\cite{lq_pairs}. In contrast, single
production takes place via an unknown $R$-parity violating Yukawa
interaction $\lambda$. Nevertheless, the available phase space for
single stop (sbottom/leptoquark) production is larger than the one for
pair production, allowing the search to extend considerably the bounds
on these particles~\cite{oscar}. Moreover, the single production can
also give information on the Yukawa couplings $\lambda$.

\smallskip

In this letter, we assume that the superpotential exhibits a term
$\lambda_{ijk}^\prime \epsilon_{ab} L^a_i Q^b_j D^c_k $ where $L$
($Q$) stands for the lepton (quark) doublet superfield and $D^c$ is
the charge conjugates right handed quark superfield. This way scalar
tops ($\tilde{t}_1$) and sbottoms ($\tilde{b}_1$) couple to
quark--lepton pairs just like a scalar leptoquark. The production of
single stops (sbottoms/leptoquarks) in association with a charged or
neutral lepton proceeds via~\cite{single_lq}:
\begin{equation}
        p \bar{p}~(pp) \to q g \to \ell \tilde{t}_1 
\end{equation}
We study this process taking into account the SUSY-QCD next-to-leading
order corrections. We show how higher order corrections not only
enhance the total cross section, but also render the theoretical
predictions more stable with respect to variations of the
renormalization and factorization scales.

\smallskip

All our results, of course, apply to models containing a single
leptoquark in addition to the Standard Model particles in the limit in
which we decouple all supersymmetric states but the lightest stop or
sbottom. This way, we can obtain the QCD corrections to the single
leptoquark production and analyze their impact on the attainable
Tevatron and LHC limits. In Section~\ref{sec:decoupling} we describe
in detail the decoupling properties of the heavy supersymmetric states
including next-to-leading order effects in linking supersymmetric with
leptoquark-type observables.

\smallskip

\underline{Conventions:} Throughout this paper we show
consistent leading order or next-to-leading order cross section
predictions, including the respective one loop or two loop strong
coupling constant and the corresponding CTEQ5L or CTEQ5M1 parton
densities~\cite{cteq5}. We usually assume a scalar top
(bottom/leptoquark) mass of $200\gev$, a massless lepton, and we set
the $R$-parity violating coupling $\lsusy$ to unity. Furthermore, we
assume that the squark couple to a quark and a lepton without any
additional suppression from the squark mixing angle. Since the
$R$-parity violating coupling as well as the mixing angle dependence
are universal factors, as far as QCD corrections are concerned, they
can be trivially added. If not stated otherwise, we scale the
supersymmetric mass spectrum as $\mg=\mse+100\gev$,
$\ms=\mse+200\gev$, and $\msz=\mse+300\gev$\footnote{The FORTRAN code
used in this letter is available from the authors:
aalves@@ift.unesp.br, tilman.plehn@@cern.ch.}.

\section{Next-to-Leading Order Cross Section}

\begin{figure}[t]
\begin{center}
\includegraphics[width=8.0cm]{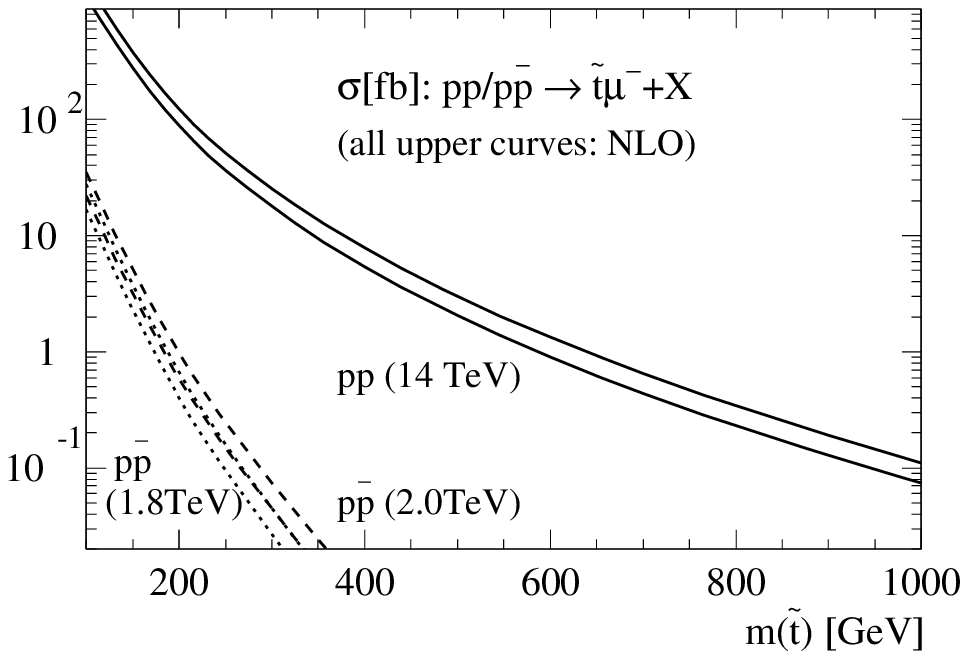} \hspace*{12mm}
\includegraphics[width=8.0cm]{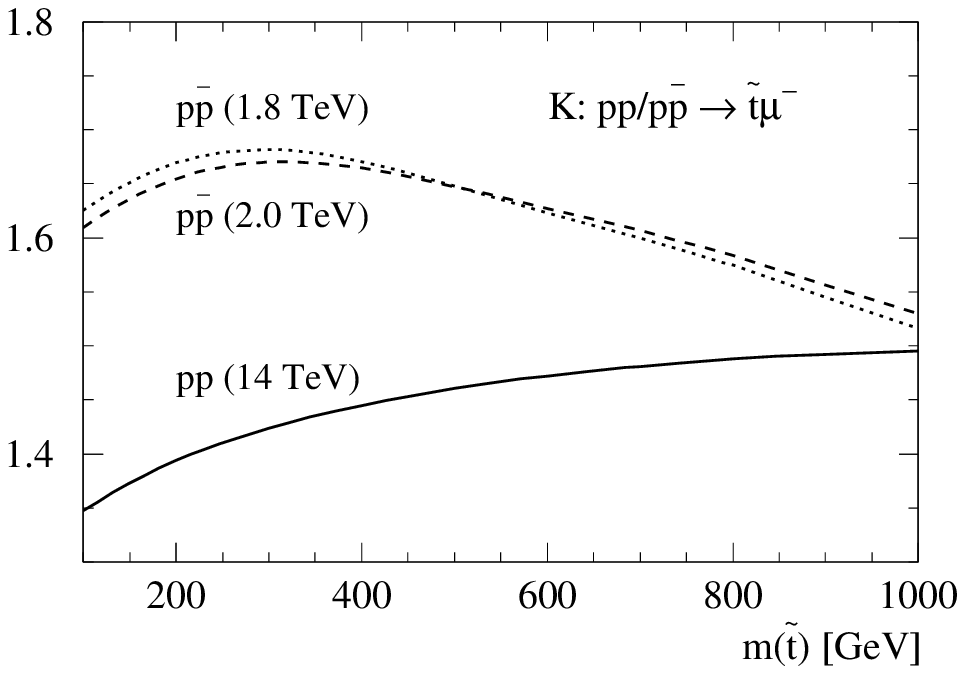} \\
\includegraphics[width=8.0cm]{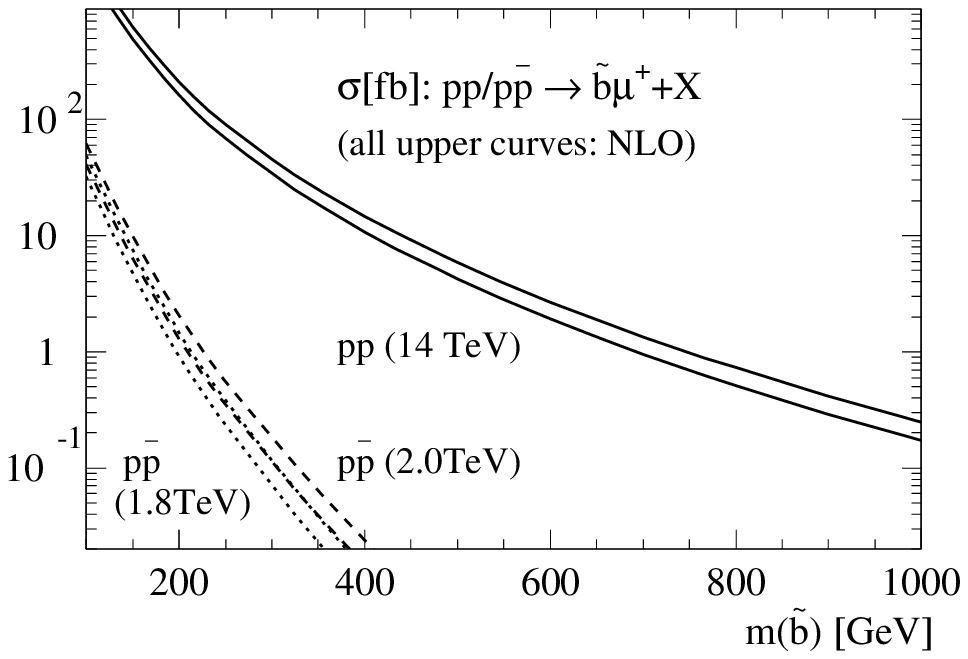} \hspace*{12mm}
\includegraphics[width=8.0cm]{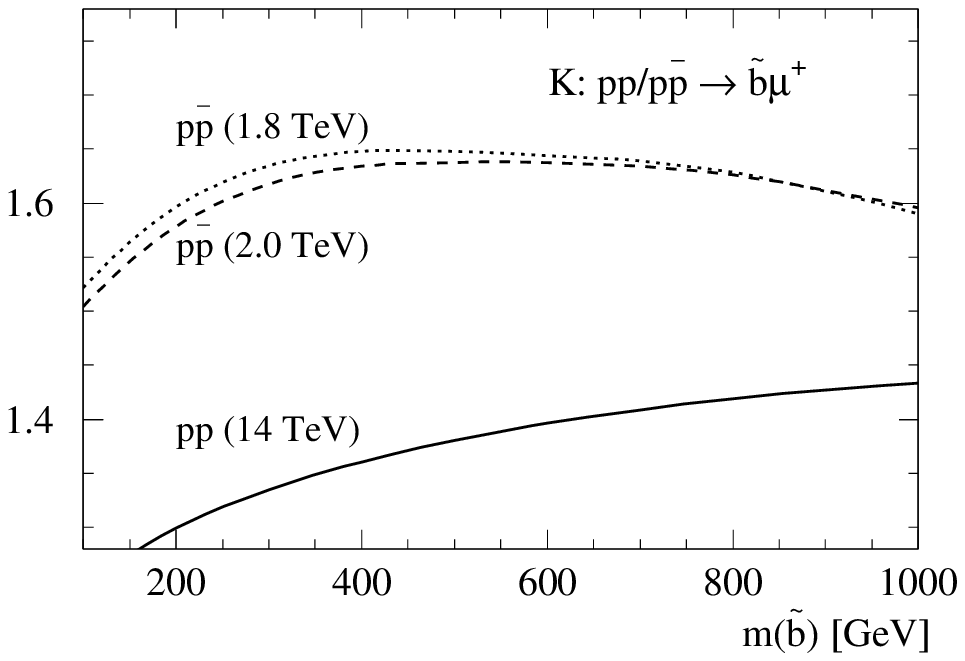} \\
\end{center}
\caption{
 The impact of the next-to-leading order corrections for the
 associated production of a top (upper row) and bottom (lower row)
 squark as a function of the particle mass, shown for the leading
 order and next-to-leading order cross sections (left) and the $K$
 factor (right). All scales are set to their central value, \ie the
 squark mass. The SUSY spectrum as a function of the stop (sbottom)
 mass is given in the text. The squark mixing angle, as well as the
 $R$ parity violating coupling $\lsusy$, are set to unity.}
\label{fig:kfac}
\end{figure}

The leading order partonic cross section for the single stop
(sbottom/leptoquark) production is given by:
\begin{equation}
 {d\hat\sigma \over d\hat t} = {\lambda'}^2_{\rm SUSY}
{\alpha_s \over 24 \hat s^2}
 \left[ - {\hat t \over 2\hat s} 
        - {\hat t \hat u^2 \over \hat s(\hat u- \mse^2)^2}
        + {\hat u \hat t\over \hat s (\hat u - \mse^2)} 
 \right] 
\end{equation}
Here $\hat{s}$ is the parton--parton center-of-mass energy, $\hat{t} =
(p_q - p_{\tilde{t}_1})^2$, and $\hat{u} = (p_q - p_\ell)^2$. From the
definition of $\lsusy$ in the Lagrangean we see that left handed stops
are produced only in association with charged leptons. In order to
avoid strong bounds on $\lsusy$ and have a clear and unsuppressed
signal, we can for example identify $\lsusy = \lambda'_{231}$, which
implies that the initial quark has down flavor and the associated
lepton is a muon. In the case of sbottoms, the Lagrangean allows their
production in association either with a charged lepton or with a
neutrino. In that case we, for example, identify $\lsusy =
\lambda^\prime_{213}$, which corresponds to an incoming up quark and
again a muon in the final state. We emphasize that all higher order
results are the same for sbottom--lepton and sbottom--neutrino
production, since the SUSY-QCD corrections do not see the charge of
the lepton.

\smallskip

The complete set of ${\cal O}( {\lambda'}^2_{\rm SUSY} \alpha^2_s)$
corrections includes virtual gluon and gluino diagrams and real gluon
emission diagrams, as well as the (crossed) processes $gg \to \ste
\ell \bar{q}$, $q \bar{q} \to \ste \ell \bar{q}$, and 
$qq \to \ste \ell q$. Several flavor combinations of the incoming and
outgoing quarks in these crossed processes are possible. The treatment
of heavy supersymmetric states is reviewed in
Section~\ref{sec:decoupling}.  We renormalize the couplings $\lsusy$
and $\alpha_s$ in the $\overline{\rm MS}$ scheme, while the final
state stop (sbottom/leptoquark) mass and the squark mixing angle are
renormalized using a generalized on--shell scheme, which includes a
running mixing angle, evaluated with the stop mass as the fixed
renormalization scale. Using this scheme for the mixing
angle~\cite{stop_angle} is known to lead to problems for weak ${\cal
O}(\alpha)$ corrections~\cite{yamada}, but it is certainly best suited
for QCD corrections. The observable cross section in terms of a
running mixing angle can, of course, be linked to the same observable
in any other renormalization scheme, and the numerical effect has been
shown to be negligible for a large set of renormalization
schemes~\cite{stop_angle}. The purely gluon or quark induced
subprocesses lead to a double counting with pair production in the
case of an intermediate on--shell stop (sbottom/leptoquark):
$gg,q\bar{q} \to \ste \ste^* \to \ste (\ell \bar{q})$. The on--shell
contributions of these processes are usually evaluated as stop pair
production with a subsequent $R$-parity violating decay. The
off--shell contributions, however, should be part of the
next-to-leading order corrections to the associated stop and lepton
production. We, therefore, split the contributions into these two
parts, using the small width approximation, and explicitly subtract
the on--shell contribution coming from the squark pair production in
the corresponding phase space points~\cite{prospino}.

\smallskip

We display in Fig.~\ref{fig:kfac} the effect of the next-to-leading
order SUSY-QCD corrections to the associated stop--lepton and
sbottom--lepton productions. While the actual value of the cross
sections depends on the numerical value of the $R$-parity violating
coupling $\lsusy$ and on the squark mixing angle, both of them drop
out for the $K$ factor, which is defined consistently as
$K=\sigma_{\rm NLO}/\sigma_{\rm LO}$. As expected, there is hardly any
difference for the Tevatron Run~I and Run~II results. Because the
limited energy of the Tevatron makes it increasingly harder to radiate
additional jets, we see that the next-to-leading order correction
becomes smaller for heavier squark masses. The $K$ factor at the LHC
looks qualitatively different. While the center-of-mass energy is
large compared to the squark mass, a sizeable fraction of the
next-to-leading order cross section comes from two initial state
gluons. For small squark masses it is likely that the purely gluonic
initial state first produces an on-shell squark pair. This
contribution we subtract, which automatically yields a small $K$
factor. For larger squark masses more of the intermediate stops will
be off--shell and, thereby, contribute to the $K$ factor. On a very
moderate level the difference between the $K$ factors for the Tevatron
Run~I and Run~II already shows the same effect which we see very
clearly at the LHC. We emphasize that the smaller $K$ factor for the
LHC, as seen in Fig.~\ref{fig:kfac} is entirely a function of the
squark mass and not a feature of the next-to-leading order corrections
at the LHC.

\smallskip

In Fig.~\ref{fig:kfac} we also observe a difference in the cross
section and in the $K$ factor between stop and sbottom production. For
the purely QCD corrections this is caused by the exchange of an
incoming down quark in the stop case by an incoming up quark in the
sbottom case. The supersymmetric corrections involving virtual gluinos
require virtual quarks in the loops which match the flavor of the
final state squark. Since the top mass is essentially of the same
order of the supersymmetric mass scale and the bottom mass is very
much smaller, this effect becomes visible. We note, however, that the
behavior of the $K$ factor as a function of the squark mass at the
Tevatron and at the LHC is the same for final state stops and
sbottoms, as one would expect from the arguments given above.

\smallskip

\begin{figure}[t]
\begin{center}
\includegraphics[width=8.0cm]{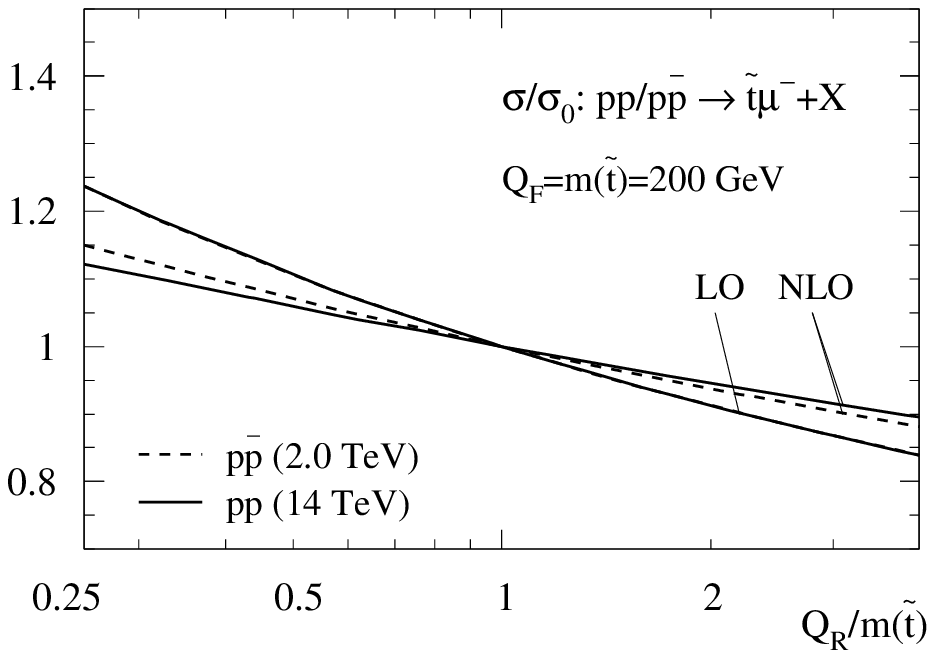} \hspace*{12mm}
\includegraphics[width=8.0cm]{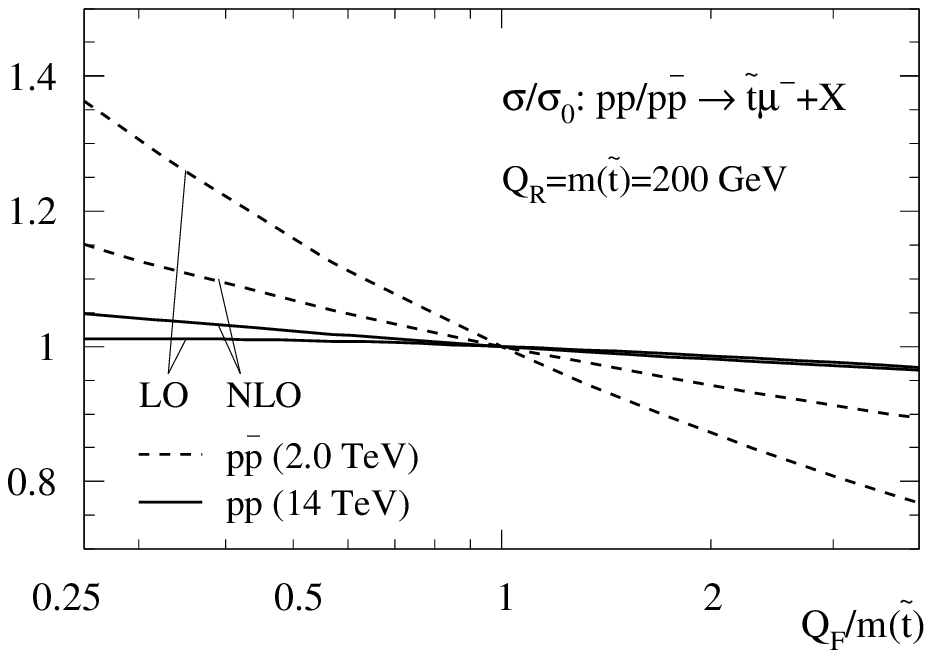}
\end{center}
\caption{ 
 Dependence of the total cross sections on the renormalization (left)
 and factorization (right) scales. The other respective scale is fixed
 to the central scale. The stop mass is set to $200\gev$ and fixes the
 central value for the scales as well as the SUSY spectrum.}
\label{fig:scale}
\end{figure}

For production cross sections involving strongly interacting final
states, the appropriate measure for the theoretical uncertainty coming
from higher order corrections is the renormalization and factorization
scale dependences. We show these scale dependences in
Fig.~\ref{fig:scale}. It is common to identify both of these scales,
however, this can lead to systematic cancellations in the variation of
the cross section with this common
scale~\cite{charged_higgs,single_susy_nlo}. This, in turn, would yield
a significant underestimate of the theoretical uncertainty.  The scale
dependence for the associated stop--lepton production is shown for a
stop mass of $200\gev$ and the corresponding SUSY spectrum.  If one
defines a theoretical error of the leading order cross section
corresponding to the variation $Q=\mse/4 \cdots 4 \mse$ of the
renormalization scale dependence, the theoretical uncertainty at the
Tevatron (LHC) is of the order of $20\%$. Notice that, in leading
order, this uncertainty comes from the variation of $\alpha_s(Q_R)$.
To next-to-leading order this error is reduced to $15\%$ for the
Tevatron, and slightly less for the LHC.

\smallskip

In contrast, the factorization scale dependence of the leading order
cross section is not identical for the Tevatron and the LHC. This is
due to the fact that at the Tevatron the stops have to be produced right
at threshold while at the LHC they can be produced through partons
with considerably lower parton momentum fraction $x$. The leading
order factorization scale dependence of this process at the LHC is
extraordinarily mild and, in comparison with other similar
processes~\cite{prospino,single_susy_nlo}, artificially flat. This is
an effect of the choice of the renormalization scale and a
cancellation involving a combination of logs $\log Q_R^2/\mse^2
\times \log Q_F^2/\mse^2$~\cite{charged_higgs}. The leading order
factorization scale dependence actually changes the sign of the slope
from positive to negative values when the stop mass increases from
$100$ to $500\gev$. Evaluating the scale dependence of the total cross
section at the LHC for a stop mass of $200\gev$ is very close to the
turnover point, at which the factorization scale is actually flat.

\smallskip

Both leading order scale uncertainties add up to some $65\%$ for the
Tevatron and $40\%$ for the LHC. The next-to-leading order
calculations reduce this uncertainty significantly to $\sim 25\%$ for
both colliders. Since there is no cancellation in the cross section
between the two scale variations at next-to-leading order, we can
obtain the same estimates by varying only the identified scale
$Q=Q_R=Q_F$. These percentages are, of course, only a rough
theoretically motivated estimate. In particular, the leading order
uncertainty is more dependent on the powers of $\alpha_s$ in the cross
section than on the actual process~\cite{prospino} and, therefore, has
to be taken with a grain of salt. A comparison with the actual $K$
factor given in Fig.~\ref{fig:kfac} also shows that the estimated
leading order uncertainty for the LHC is indeed too small, while the
leading order error band covers the next-to-leading order curve well
for the Tevatron. Furthermore, to next-to-leading order different
supersymmetric production processes give very consistent uncertainty
estimates~\cite{single_susy_nlo,lq_pairs,prospino,charged_higgs}. Since
there is no physics reason why the Tevatron and the LHC cross section
should behave any differently as far as the theoretical uncertainty is
concerned, it is a good check to see that at the next-to-leading order
level the scale dependence is indeed similar for both
experiments. Moreover, the next-to-leading error bands agrees well
with what one would expect from similar processes.\bigskip

\smallskip

\begin{figure}[t]
\begin{center}
\includegraphics[width=8.0cm]{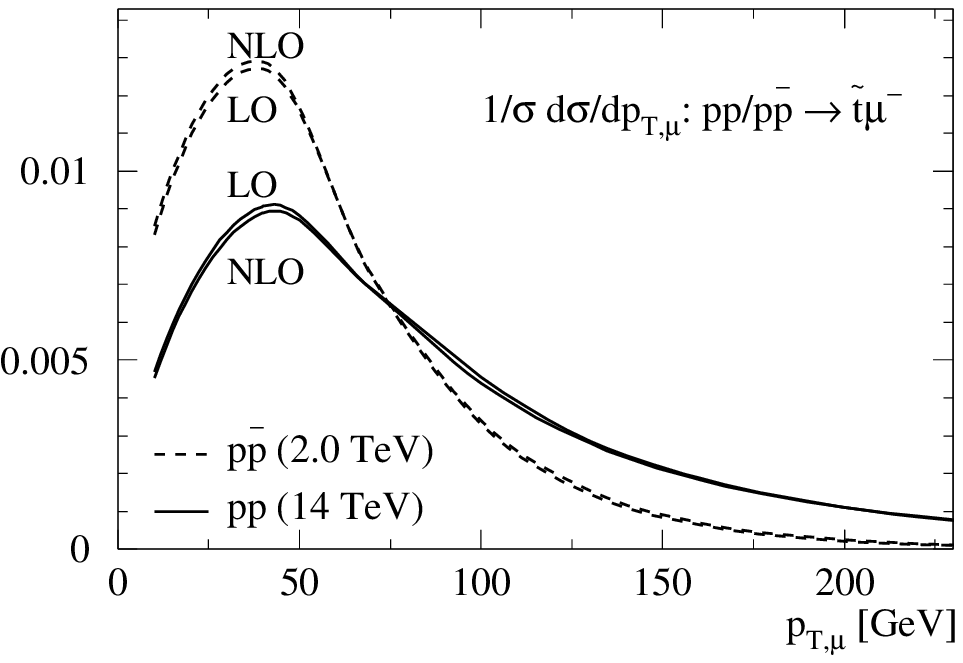} \hspace*{12mm}
\includegraphics[width=8.0cm]{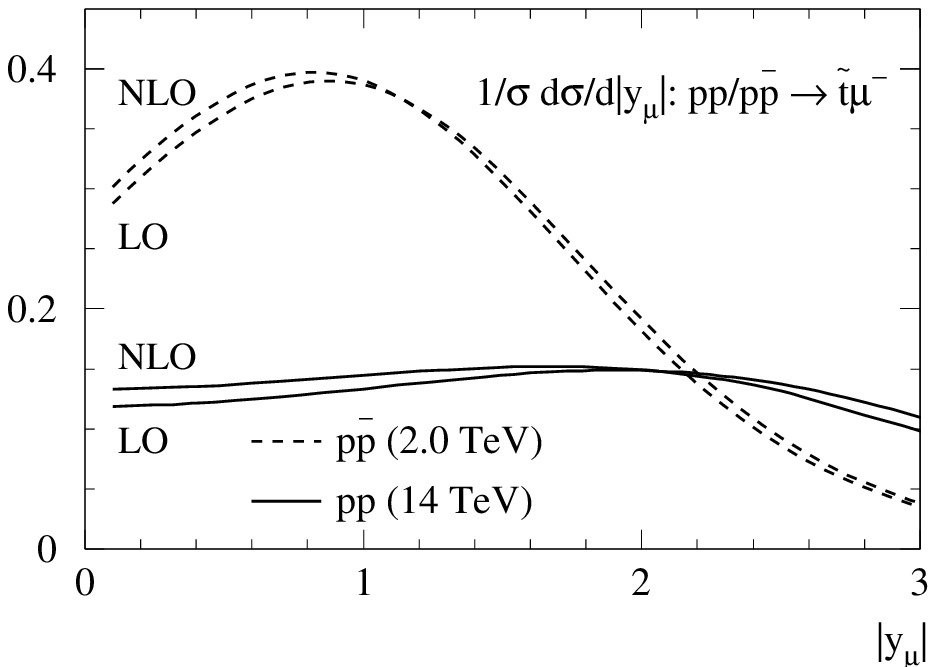}
\end{center}
\caption{ 
 The normalized charged lepton transverse momentum (left) and rapidity
 (right) distributions at the Tevatron and at the LHC. The stop mass
 is set to $200\gev$ and fixes the SUSY spectrum as well as the
 scale.}
\label{fig:distri}
\end{figure}

We present in Fig.~\ref{fig:distri} the effect of the next-to-leading
order contributions on the lepton transverse momentum and rapidity
distributions. We generally observe that the next-to-leading order
corrections change the shape of the lepton transverse momentum and
rapidity distributions only at a negligible level. Looking into the
transverse momentum distributions in more detail, we see that, while
for the Tevatron the lepton is expected to be slightly softer after
including the next-to-leading order corrections, at the LHC the lepton
tends to be harder. The reason is again the ratio of the stop mass and
the collider energy: at the Tevatron all stops are produced very close
to threshold with little transverse momentum. Therefore, there is very
little energy available to produce additional hard jets; soft and
collinear radiation will dominate the next-to-leading order effects
and reduce the available partonic energy for the hard process. This
renders the event softer altogether. In contrast, at the LHC there is
more energy available to actually produce hard jets in initial and
final state radiation. While initial state radiation still lowers the
energy available for the hard process, final state radiation off the
stop has to be balanced by the lepton. These two effects cancel to a
very large degree but yield a slightly harder lepton spectrum. Because
the leading order and next-to-leading order distributions are very
similar, it is a sufficient approximation to rescale the leading order
distributions by the constant $K$ factor obtained from the total cross
section.

\smallskip

The salient feature of the right panel of Fig.~\ref{fig:distri} is
that the lepton rapidity distribution does not have its maximum at
zero rapidity. In other words, the lepton prefers to be
boosted. Looking at the initial state, we see that for the Born
process the two incoming partons are one gluon and one valence
quark. On average the latter will have the larger parton energy. This
will boost the event altogether into the direction of the valence
quark, which in turn, boosts the lepton into the forward direction. As
can be seen in Fig.~\ref{fig:distri}, the lepton is even farther
forward at the LHC than it is at the Tevatron, which happens because
at the LHC a larger fraction of events actually probes smaller parton
momentum fractions $x$ of the gluon and leads to larger momentum
imbalance of the initial state. The next-to-leading order prediction
softens the bias toward forward leptons at both colliders because of
the final state radiation yields another hard central jet radiated
from the stop, which has to be balanced by the lepton.

\section{Decoupling of Heavy Supersymmetric Particles}
\label{sec:decoupling}

While most of the calculation described in this paper has been done in
the framework of $R$-parity violating supersymmetry, it is well known
that the same kind of signal can be generated simply extending the
Standard Model by a single scalar leptoquark \cite{lq_theory}. This
scalar leptoquark would correspond in supersymmetry to the lightest
squark, which can be either the lighter stop ($\tilde{t}_1$) or the
lighter sbottom ($\tilde{b}_1$). The $R$-parity violating coupling
$\lsusy$ corresponds to the scalar leptoquark coupling $\llq$. In
other words, if we decouple all additional strongly interacting
supersymmetric particles, we should recover the Standard Model with an
additional scalar leptoquark in leading order as well as in
next-to-leading order QCD.\smallskip

The decoupling of heavy strongly interacting states is slightly
complicated by the contribution of these states to the running of
$\msbar$ quantities and their identification with the corresponding
Standard Model quantities.  As an example we recall the treatment of
heavy states in the evolution of the strong coupling $\partial
\alpha_s/\partial \log \mu_R^2 = - \beta_{\alpha} \alpha_s / (4\pi)$.  
The current extraction of the strong coupling from data explicitly
uses only five quark flavors, while the renormalization of the same
quantity in the Standard Model involves all six quarks. As a matter of
fact, the $\msbar$ $\alpha_s$ renormalization in the MSSM leads to the
complete one-loop beta function, which does not exhibit the decoupling
of the heavy states at low energy
\begin{equation}
\beta_{\alpha} = \beta_{\alpha}^{\rm SM} + \beta_{\alpha}^{\rm SUSY} 
              = \left( \frac{11}{3}N - \frac{2}{3} n_f \right) \; + \;
                \left(-\frac{2}{3}N - \frac{1}{3} n_s \right) 
\end{equation}
The supersymmetric contribution comes from gluino and from squark
loops; in the MSSM the number of quark and squark flavors is
$n_f=n_s=6$. To match the measured $\alpha_s$, we have to explicitly
decouple the heavy particles from the running of $\alpha_s$, {\em
i.e.} we have to include an additional term in the next-to-leading
order corrections~\cite{prospino}:
\begin{equation}
\frac{\Delta \alpha_s}{\alpha_s} = \frac{\alpha_s}{4\pi} \; \left[
            \frac{1}{6} \log \frac{\mu_R^2}{\mt^2}
          + \frac{n_f-1}{3} \log \frac{\mu_R^2}{\ms^2}
          + \frac{1}{6} \log \frac{\mu_R^2}{\mse^2}
          + \frac{1}{6} \log \frac{\mu_R^2}{\msz^2}
          + \frac{N}{6} \log \frac{\mu_R^2}{\mg^2}
            \right] 
\end{equation}
This contribution explicitly cancels the contribution of the heavy particles
to the running of $\alpha_s$ and ensures that the one-loop renormalization of
$\alpha_s$ is governed by $\beta_{\alpha}^{\rm SM}$ with $n_f=5$.\smallskip

\begin{figure}[t]
\begin{center}
\includegraphics[width=8.0cm]{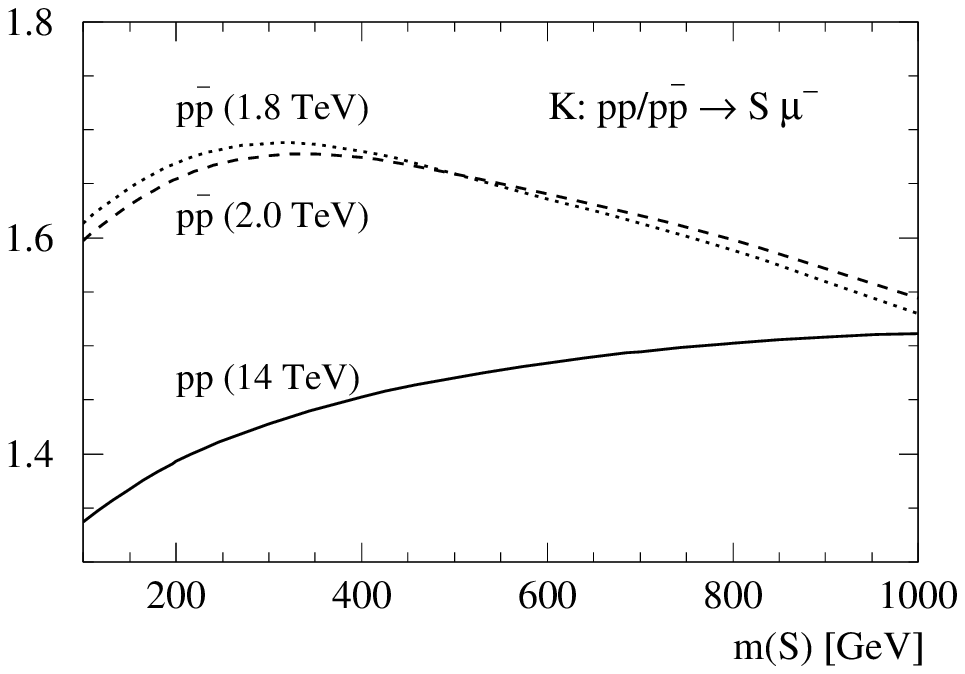} \hspace*{12mm} 
\includegraphics[width=8.0cm]{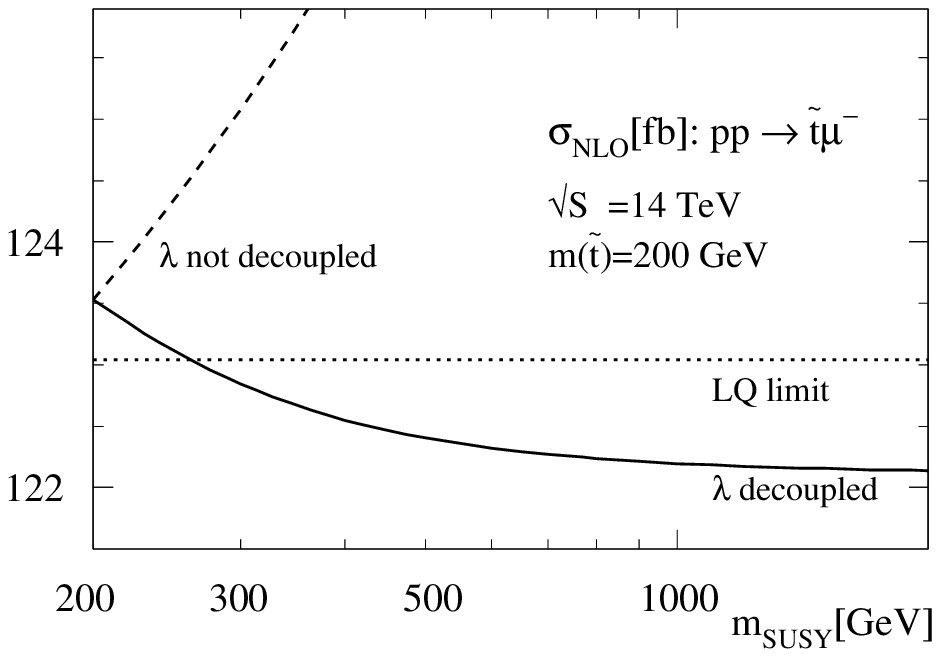}
\end{center}
\caption{ 
 Left: the impact of the next-to-leading order QCD corrections on the
 leptoquark production cross sections. All input parameters are
 identical to the right panel of Fig.~\ref{fig:kfac}, the leptoquark
 coupling is derived from the stop case, \ie the scalar leptoquark
 couples to an incoming down quark and an outgoing muon. Right: the
 decoupling behavior of the supersymmetric single stop production
 cross section at next-to-leading order. The dashed curve we compute
 without decoupling the heavy supersymmetric particles from the
 running of the $R$ parity violating coupling $\lsusy$, as described
 in Eq.\ (\ref{eq:dec2}).}
\label{fig:decoup}
\end{figure}

As expected, we find a completely analogous behavior for the running of the
$R$-parity violating coupling: $\partial \lsusy/\partial \log \mu_R^2 = -
\beta_{\lambda}\alpha_s/(4\pi) $. At the one-loop level it turns out that the
beta function of $\lsusy$ vanishes. This is not a generic feature of
the MSSM but a pure accident. For example, the $R$-parity violating
coupling $\lambda''$ has a finite one-loop beta function in the MSSM
as well as in the limit of heavy gluinos~\cite{single_susy_nlo}. The
beta function of $\llq$ is well-known~\cite{hera_nlo}:
\begin{equation}
\beta_{\lambda} = \beta_{\lambda}^{\rm LQ} + \beta_{\lambda}^{\rm SUSY}
                = \left( \frac{3}{2} C_F \right) \; + \; 
                  \left(-\frac{3}{2} C_F \right) 
\end{equation}
Since neither $\llq$ nor $\lsusy$ are measured parameters (yet) we can
in principle choose any definition as long as the calculation is
consistent.  However, we cannot use the constant coupling $\lsusy$ and
naively decouple the heavy supersymmetric particles to obtain the
scalar leptoquark theory: the naive supersymmetric corrections would
be divergent with $\log \mheavy^2$.  Again we have to include a
one-loop decoupling term to cancel this logarithmic divergence of the
next-to-leading order cross section. Throughout the numerical analysis
we include this divergent next-to-leading order decoupling
contribution.  Nevertheless, we also find that in the $\msbar$ scheme
we are left with a finite threshold correction difference between
$\lsusy$ and $\llq$. A complete decoupling term would read:
\begin{equation}
\frac{\Delta \lsusy}{\lsusy} = \frac{1}{2} \; \frac{\alpha_s}{4\pi} \; \left[
             3 \, C_F \log \frac{\mu_R^2}{\mg^2}
           + C_F
             \right] \; .
\label{eq:dec2}
\end{equation}\smallskip

Notice that, in contrast to all divergent decoupling corrections, this
last term does not vanish in the case of a light supersymmetric
spectrum $\mu_R \sim \mse \sim \mg$. For this reason we do {\sl not}
include it in our numerical analyses. Initially, we can see in the
left panel of Fig.\ \ref{fig:decoup} that the next-to-leading order
corrections for a stop--type leptoquark coupling to an initial state
down quark behave exactly like the ones for supersymmetric stops. In
the right panel we exhibit the explicit decoupling of the heavy
supersymmetric state, for which we assume the gluino mass,
light-flavor squark mass, and the heavier stop mass as being
degenerate and equal to $M_{\text{SUSY}}$.  It is interesting to
notice the extremely large effect that takes place if you do not
decouple $\lambda$.  Once all issues described above are taken care
of, the supersymmetric cross section approaches a decoupling
limit. However, the finite difference given in Eq.\ (\ref{eq:dec2})
remains between the supersymmetric decoupling limit and the leptoquark
next-to-leading order model.

\smallskip

For the sake of completeness, we would like to point out that a
consistent treatment of heavy flavors is also required in other
sectors of the Standard Model. The top and bottom Yukawa couplings are
running $\msbar$ parameters with contributions from heavy
supersymmetric particles. These have to be removed from the running in
the limit of a heavy supersymmetric spectrum exactly the way we
described it for $\alpha_s$ and $\lsusy$~\cite{charged_higgs}. In
contrast, the stop mass and the stop mixing angle in this paper are
defined in the on--shell scheme, which does not involve any running
and which is based on an independent physics condition. Therefore,
they do not pose any problems for a decoupled supersymmetric spectrum.

\section{Discussion and Outlook}

The single production of stops (sbottoms/leptoquarks) has the
potential of extending the reach of colliders to discover or rule out
their existence~\cite{single_slepton,single_stop,single_susy_nlo}. As
we have shown, the inclusion of higher order SUSY-QCD corrections
renders the theoretical predictions for these processes more stable
with respect to variations of the renormalization and factorization
scales. Moreover, these corrections enhance the total cross section by
a factor up to $70$\% at the Tevatron and 50\% at the LHC. This leads
to a larger sensitivity to stops (sbottoms/leptoquarks). 

\smallskip

To illustrate the effect of QCD next-leading-order corrections on the
searches for new physics, let us consider the single production of
leptoquarks. Using the Born expressions for the cross sections, the
attainable limits on leptoquarks coupling to up (down) quarks and
charged leptons is 310 (260) GeV and 2.9 (2.4) TeV for the Tevatron
RUN II and LHC, respectively~\cite{oscar}, assuming that the
leptoquark decays exclusively into a charged lepton and a
quark. Taking into account the $K$ factors displayed in Fig.\
\ref{fig:decoup} (left panel) these limits read 350 (280) GeV and 3.1
(2.6) TeV for the RUN II and LHC respectively.

\smallskip

It is interesting to notice that the single production of stops or
sbottoms give us the opportunity to direct measure their Yukawa
coupling to lepton--quark pairs, which otherwise would only be
possible indirectly through their effects on the Drell-Yan
process~\cite{rohini}.


\section*{Acknowledgements}

O.E. would like to thank the Theory Division of CERN for the
hospitality in the final stage of this work.  This work was supported
in part by Conselho Nacional de Desenvolvimento Cient\'{\i}fico e
Tecnol\'ogico (CNPq), by Funda\c{c}\~ao de Amparo \`a Pesquisa do
Estado de S\~ao Paulo (FAPESP), and by Programa de Apoio a N\'ucleos
de Excel\^encia (PRONEX), by DOE grant DE-FG02-95ER-40896, and the
University of Wisconsin Research Committee with funds granted by the
Wisconsin Alumni Research Foundation.



\begin{thebibliography}{99}

\bibitem{expl}
 G.~G.~Ross and J.~W.~Valle,
  Phys.\ Lett.\ B {\bf 151}, 375 (1985);
 S.~Dimopoulos and L.~J.~Hall,
  Phys.\ Lett.\ B {\bf 207}, 210 (1988);
 V.~D.~Barger, G.~F.~Giudice and T.~Han,
  Phys.\ Rev.\ D {\bf 40}, 2987 (1989);
 E.~Ma and D.~Ng,
  Phys.\ Rev.\ D {\bf 41}, 1005 (1990);
 T.~Banks, Y.~Grossman, E.~Nardi and Y.~Nir,
  Phys.\ Rev.\ D {\bf 52}, 5319 (1995).

\bibitem{rp_coll}
 H.~K.~Dreiner and G.~G.~Ross,
  Nucl.\ Phys.\ B {\bf 365}, 597 (1991);
 J.~Kalinowski, R.~R\"uckl, H.~Spiesberger and P.~M.~Zerwas,
  Phys.\ Lett.\ B {\bf 414}, 297 (1997);
 B.~Allanach {\it et al.},
  arXiv:hep-ph/9906224;
 P.~Richardson,
  arXiv:hep-ph/0101105.

\bibitem{single_slepton}
 H.~K.~Dreiner, P.~Richardson and M.~H.~Seymour,
  arXiv:hep-ph/9903419
  and 
  arXiv:hep-ph/0001224;
 G.~Moreau, M.~Chemtob, F.~Deliot, C.~Royon and E.~Perez,
  Phys.\ Lett.\ B {\bf 475}, 184 (2000);
 G.~Moreau, E.~Perez and G.~Polesello,
  Nucl.\ Phys.\ B {\bf 604}, 3 (2001).

\bibitem{single_stop}
 E.~L.~Berger, B.~W.~Harris and Z.~Sullivan,
  Phys.\ Rev.\ Lett.\  {\bf 83}, 4472 (1999)
 and 
  Phys.\ Rev.\ D {\bf 63}, 115001 (2001)

\bibitem{single_susy_nlo}
 T.~Plehn,
  Phys.\ Lett.\ B {\bf 488}, 359 (2000);
 D.~Choudhury, S.~Majhi and V.~Ravindran,
  arXiv:hep-ph/0207247.

\bibitem{lq_pairs}
 B.~Dion, L.~Marleau and G.~Simon,
  Phys.\ Rev.\ D {\bf 56}, 479 (1997);
 M.~Kr\"amer, T.~Plehn, M.~Spira and P.~M.~Zerwas,
  Phys.\ Rev.\ Lett.\  {\bf 79}, 341 (1997).

\bibitem{oscar} 
 This has been shown for leptoquarks: 
 O.~J.~\'Eboli, R.~Zukanovich Funchal and T.~L.~Lungov,
  Phys.\ Rev.\ D {\bf 57}, 1715 (1998);
 O.~J.~\'Eboli and T.~L.~Lungov,
  Phys.\ Rev.\ D {\bf 61}, 075015 (2000).

\bibitem{single_lq} 
 J.~L.~Hewett and S.~Pakvasa,
  Phys.\ Rev.\ D {\bf 37}, 3165 (1988);
 O.~J.~\'Eboli and A.~V.~Olinto,
  Phys.\ Rev.\ D {\bf 38}, 3461 (1988).

\bibitem{cteq5}
 H.~L.~Lai {\it et al.}  [CTEQ Collaboration],
  Eur.\ Phys.\ J.\ C {\bf 12}, 375 (2000).

\bibitem{stop_angle}
 W.~Beenakker, R.~H\"opker, T.~Plehn and P.~M.~Zerwas,
  Z.\ Phys.\ C {\bf 75}, 349 (1997);
 T.~Plehn,
  arXiv:hep-ph/9809319;
 for a comparison with different renormalization schemes see \eg 
 S.~Kraml, H.~Eberl, A.~Bartl, W.~Majerotto and W.~Porod,
  Phys.\ Lett.\ B {\bf 386}, 175 (1996);
 A.~Djouadi, W.~Hollik and C.~J\"unger,
  Phys.\ Rev.\ D {\bf 55}, 6975 (1997).

\bibitem{yamada}
 Y.~Yamada,
  Phys.\ Rev.\ D {\bf 64}, 036008 (2001);
 W.~Hollik, E.~Kraus, M.~Roth, C.~Rupp, K.~Sibold and D.~Stockinger,
  Nucl.\ Phys.\ B {\bf 639}, 3 (2002).

\bibitem{prospino}
 W.~Beenakker, R.~H\"opker, M.~Spira and P.~M.~Zerwas,
  Nucl.\ Phys.\ B {\bf 492}, 51 (1997);
 W.~Beenakker, M.~Kr\"amer, T.~Plehn, M.~Spira and P.~M.~Zerwas,
  Nucl.\ Phys.\ B {\bf 515}, 3 (1998);
 W.~Beenakker, M.~Klasen, M.~Kr\"amer, T.~Plehn, M.~Spira and P.~M.~Zerwas,
  Phys.\ Rev.\ Lett.\  {\bf 83}, 3780 (1999).

\bibitem{charged_higgs}
 T.~Plehn,
  Phys.\ Rev.\ D in print, 
  arXiv:hep-ph/0206121.
 
\bibitem{lq_theory}
 W.~Buchm\"uller and D.~Wyler,
  Nucl.\ Phys.\ B {\bf 268}, 621 (1986);
 W.~Buchm\"uller, R.~R\"uckl and D.~Wyler, 
  Phys.\ Lett.\ B {\bf 191}, 442 (1987)
  [Erratum-ibid.\ B {\bf 448}, 320 (1999)].


\bibitem{hera_nlo}
 Z.~Kunszt and W.~J.~Stirling,
  Z.\ Phys.\ C {\bf 75}, 453 (1997);
 T.~Plehn, H.~Spiesberger, M.~Spira and P.~M.~Zerwas,
  Z.\ Phys.\ C {\bf 74}, 611 (1997).

\bibitem{rohini}
 D.~Choudhury, R.~M.~Godbole and G.~Polesello,
  JHEP {\bf 0208} (2002) 004.

\end{thebibliography}
\end{document}